\documentclass{article}

\usepackage[ margin=15mm]{geometry}

\usepackage{authblk}
\usepackage{graphicx}
\usepackage{caption}
\usepackage{amsmath,amssymb,amsfonts,amsthm}
\usepackage{newtxmath}
\usepackage{mathrsfs}
\usepackage{hyperref}
\usepackage[capitalise]{cleveref}
\usepackage{natbib}

\newlength{\figwidth}
\theoremstyle{thmstyleone}
\newtheorem{theorem}{Theorem}
\newtheorem{proposition}[theorem]{Proposition}
\newtheorem{lemma}[theorem]{Lemma}
\newtheorem{corollary}[theorem]{Corollary}

\theoremstyle{thmstyletwo}
\newtheorem{example}{Example}
\newtheorem{remark}{Remark}

\theoremstyle{thmstylethree}
\newtheorem{definition}{Definition}

\renewcommand{\P}{\ensuremath{\mathcal{P}}} 
\newcommand{\B}{\ensuremath{\mathbb{B}}}  
\newcommand{\T}{\ensuremath{\tau}}        
\newcommand{\N}{\mathbb{N}}
\newcommand{\M}{\aleph}         
\newcommand{\I}{\mathcal{I}}    

\newcommand{\Target}{T}         
\newcommand{\Reg}{R}            

\newcommand{\Same}{\nabla}      
\newcommand{\Free}{\Psi}        
\newcommand{\EDelta}{\Delta^+}  
\newcommand{\Required}{\overrightarrow{\Delta}}     
\newcommand{\ERequired}{\overrightarrow{\Delta^+}}  

\newcommand{\mvtobool}{\Omega}        
\newcommand{\mvtobuf}{\Gamma}         

\newcommand{\extension}{\mathcal{E}}  
\newcommand{\E}{\epsilon}             
\newcommand{\refinement}{\mathcal{R}} 
\newcommand{\booltostr}{\rho}         
\newcommand{\ind}{\vmathbb{1}}        

\providecommand{\norm}[1]{{\lvert#1\rvert}}

\begin{document}

\title{Linear cuts in Boolean networks}

\author[1]{Aur\'elien Naldi}
\author[2]{Adrien Richard}
\author[3]{Elisa Tonello}

\affil[1]{Lifeware Group, Inria Saclay-Ile de France, Palaiseau, France}
\affil[2]{I3S, Université Côte d'azur - CNRS, Sophia Antipolis, France}
\affil[3]{Department of Mathematics and Computer Science, Freie Universität Berlin, Germany}


\maketitle

\abstract{
Boolean networks are popular tools for the exploration of qualitative dynamical properties of biological systems.
Several dynamical interpretations have been proposed based on the same logical structure that captures the interactions
between Boolean components. They reproduce, in different degrees, the behaviours emerging in more quantitative models.
In particular, regulatory conflicts can prevent the standard asynchronous dynamics from reproducing some trajectories
that might be expected upon inspection of more detailed models.
We introduce and study the class of networks with linear cuts, where linear components --
intermediates with a single regulator and a single target -- eliminate the aforementioned regulatory conflicts.
The interaction graph of a Boolean network admits a linear cut when a linear component occurs in each cycle and
in each path from components with multiple targets to components with multiple regulators.
Under this structural condition the attractors are in one-to-one correspondence
with the minimal trap spaces, and the reachability of attractors can also be easily characterized.
Linear cuts provide the base for a new interpretation of the Boolean semantics that captures all behaviours of
multi-valued refinements with regulatory thresholds that are uniquely defined for each interaction,
and contribute a new approach for the investigation of behaviour of logical models.
}

\section{Introduction}

Boolean networks are a class of non-deterministic discrete event systems used as qualitative
dynamical models of biological processes.
The study of complex biological processes leads to two types of results: insight about the internal molecular mechanisms, 
and observation of their state over time and different external stimulations. While the changes of state emerge from the
internal mechanisms, they can not be directly compared. The integration of mechanistic knowledge into dynamical models
enables to contrast the behaviour emerging from the model with the experimental observations.
Such models are valuable tools to identify inconsistencies, evaluate hypothesis and prioritize their experimental validation.
Starting with a known initial condition, the model can be used to predict the reachability and stability of a target phenotype,
which corresponds to properties of the reachable states of the model.
The lack of precise information on the initial conditions and kinetic parameters impedes the construction of comprehensive quantitative
models without performing time-consuming exploration of parameters.
Boolean and more generally qualitative models have been proposed to cope with this lack of quantitative knowledge~\citep{kauffman1969,thomas1973}.
These models provide a discrete approximation well suited to build large comprehensive models based on incomplete knowledge. They are also amenable to formal
analysis, in particular for the identification of attractors~\citep{Naldi2007,Dubrova2011,klarner2014trapspaces}.
Multi-valued networks can be used to account for components for which a higher activity level (denoting for example a higher concentration or
a stronger activation) can lead to different effects (new targets, stronger or different effect). Most large networks lack this level of detail and
consider only Boolean components (sometimes a few selected multi-valued components). In practice, the coarse-grained predictions obtained with these
models are sufficient to reproduce relevant behaviours in a wide range of biological applications \citep[e.g.][]{Sizek2019,Beal2021,Bonzanni2013,Cohen2015,Collombet2017}.

The analysis of these models often aims initially at the identification of attractors
(fixed points or stable oscillations) and reachability properties, which are computationally hard problems in the classical
asynchronous semantics. Modelers can attempt to simplify the analysis by first considering \emph{trap spaces}, stable subspaces
that can be efficiently identified using constraint-solving approaches~\citep{klarner2014trapspaces}. Trap spaces provide a crude approximation
of some of the attractors, but may not capture all of them. In addition, they can be used to rule out some reachability properties, since all states outside of the
smallest trap space including the initial state are not reachable. On the other hand, reachability analysis inside a given trap space remains hard to solve.
These questions are much easier to tackle using the recently proposed \emph{most permissive} semantics~\citep{pauleve2020most},
an over-approximation of the asynchronous semantics which lifts competition between concurrent events
by introducing intermediate states representing the inherent uncertainty of Boolean networks.
This approach formally accounts for the reachability properties of all possible refinements and uncovers missing realistic behaviours that are not captured by the asynchronous semantics.
This results in very good computational properties, with all attractors being trap spaces and reachability analysis being polynomial.
On the other hand, the most permissive semantics can also introduce non-monotonic behaviours, which may contradict the original intent of the model and could be considered as artefacts.

In this work, we propose an alternative semantics based on structural properties underlying the competition between components.
We study constraints on the order of events in the asynchronous and most permissive semantics, specifically those related to the existence of maximal geodesics.
We introduce the class of Boolean networks that admit a \emph{linear cut}, that is, networks in which every cycle and every path from a component with multiple targets to a component with multiple regulators contains at least one linear component (a component with in- and out-degree equal to one).
In essence, these linear components can be used to relax competitions between other components in the network.
We show that for all initial states with stable linear components (\emph{canonical states}), maximal geodesics of the most permissive semantics exist in the asynchronous dynamics.
We prove two main consequences of this observation: 1) minimal trap spaces provide a precise (but not exact)
characterization of all attractors; and 2) given a canonical initial condition, all trap spaces (hence all attractors) included in the smallest trap space containing the initial state 
are reachable. The characterization of reachability from other states and subspaces remains a hard problem. While many realistic Boolean networks do not satisfy the
required topological properties, we show that one can always construct an extended network which does, by adding intermediate components on competing interactions.
We use this extension to define a new semantics which is an over-approximation of the classical asynchronous semantics and an under-approximation of the
most-permissive semantics. In Boolean networks of biological systems, interactions are often abstract representations summarizing multiple intermediate steps,
hence networks resulting from the addition of explicit intermediates can presumably be considered as valid candidate models. In these cases, the extended semantics takes advantage of some key computational properties of the most permissive semantics with a higher confidence in the interpretability of the results.

In \cref{sec:background}, we present classical concepts and formal notation used in this work.
In \cref{sec:partial-orders}, we introduce \emph{implicant maps} as a tool to study constraints between transitions in asynchronous and permissive trajectories.
In \cref{sec:cuttable-networks}, we define the topological class of \emph{$L$-cuttable} Boolean networks and derive some of their key dynamical properties, in particular the one-to-one correspondence between minimal trap spaces and attractors.
In \cref{sec:semantics} we show that extended networks, accounting for realistic delay effects, can be used to take advantage of the dynamical properties of cuttable
networks to investigate any Boolean network, and to recreate behaviours of a class of monotonic multi-valued refinements.
Finally, we discuss how the semantics of linearly extended networks relate to the asynchronous and permissive semantics, and their potential practical application to the exploration and validation of biological models.

\section{Background}
\label{sec:background}

In this section we introduce notations and definitions used throughout the paper.
The symbol $\B$ will denote the set $\{0,1\}$.
Given a set $A$, we will write $\P(A)$ for the power set of $A$.

A \emph{Boolean network} is defined by a pair $M=(V,f)$, where $V=\{1,\dots,n\}$ is called the set of variables or components of the Boolean network, and $f$ is an endomorphism of $\B^V$.

The set $\B^V$ will be called the set of states of the Boolean network, sometimes called state space.
Any pair of states $x$ and $y$ delimit a \emph{subspace} $[x,y]$ defined as the subset of states $\{z \in \B^V\ \mid \ z_i=x_i=y_i \text{ for all } i \in V \text{ s.t. }x_i=y_i\}$.
For a subset $A$ of $\B^V$, $[A]$ will denote the minimal subspace containing $A$.
We will denote subspaces also as elements of $\{0,1,\star\}^V$, so that a state $x$ belongs to a subspace $t \in \{0,1,\star\}^V$ if for all variables $i$ we have
either $x_i = t_i$ or $t_i = \star$. That is, we use $\star$ to represent free variables.
Note that states are subspaces without free variables and that a subspace with $k$ free variables contains $2^k$ different states.

Given a subset $A$ of $\B^V$, we will denote by $\Delta(A)$ the set of components that vary in the set: $\Delta(A)=\{i \in V \mid \exists x,y \in A \text{ s.t. } x_i \neq y_i\}$.
If $A$ consists of two states $x$ and $y$, we will write $\Delta(x,y)$ for $\Delta(A)$. We extend the notation $\Delta(x,y)$ to apply to elements $x,y$ of $\{0,1,\star\}^V$.

Given a state $x \in \B^V$ and a set of components $I \subseteq V$, we define the state $\bar{x}^I$ by $\bar{x}^I_i \neq x_i$ 
for all $i \in I$ and $\bar{x}^i_j = x_j$ for all $j \in V\setminus I$.
By convention, $\bar{x}^i = \bar{x}^{\{i\}}$.

For a Boolean network $(V,f)$, we say that $i\in V$ is a \emph{regulator} of $j\in V$ if there exists $x \in \B^V$ such that $f_j(x) \neq f_j(\bar{x}^i)$. In this case, component $j$ is called a \emph{target} for $i$.
We will use the notations $\Reg,\Target \colon V \rightarrow \P(V)$ to denote the maps that give the set of regulators and targets of components, respectively.

The \emph{interaction graph} of a Boolean network $(V,f)$ summarises the regulations between components. It is the graph with set of vertices $V$ and set of edges defined by $\{(i,j) \in V \mid j \in \Target(i)\}$.
The edges of the interaction graph are also called interactions of the network.

The dynamical behaviour of a Boolean network $(V,f)$ is encoded in transitions between states.
These transitions are defined by the Boolean rules $f$ and an updating semantic, which can be deterministic (each state
has a single successor) or non-deterministic (each state can have multiple successors defining alternative
dynamical trajectories).
The deterministic synchronous updating was first proposed by \citet{kauffman1969}.
In this work, we focus on the non-deterministic asynchronous updating, introduced by \citet{thomas1973}.
As the name suggests, the synchronous updating assumes that all possible changes always happen at the same time,
while the asynchronous updating assumes that all changes happen separately.
In the generalized asynchronous updating, changes can happen either at the same time or separately: it contains
all transitions from the synchronous and asynchronous updatings, as well as all other transitions where a subset
of components are updated.
More in detail, for a Boolean network $(V,f)$, given two distinct states $x,y$ (i.e. $\Delta(x,y) \neq \emptyset$), there exists a transition from $x$ to $y$
\begin{itemize}
  \item in the synchronous dynamics, if and only if $y = f(x)$,
  \item in the asynchronous dynamics, if and only if $y = \bar{x}^i$ with $i \in \Delta(x, f(x))$,
  \item in the generalized asynchronous dynamics, if and only if $\Delta(x,y) \subseteq \Delta(x,f(x))$.
\end{itemize}
Note that each state has at most one successor in the synchronous updating, at most $n$ successors in the asynchronous updating and up to $2^n-1$ successors in the generalized asynchronous case.

Other updatings have been proposed, in particular the bloc-sequential updating \citep[deterministic, see][]{Robert1986} and the use of priority classes \citep[non-deterministic, see][]{faure2006cellcycle}.
In addition, one can define stochastic dynamics by adding transition probabilities to non-deterministic updatings.
A trajectory from a state $x$ to a state $y$ in any of these updatings implies the existence of a trajectory from $x$ to $y$ in the generalized asynchronous dynamics.
By definition, all transitions in the synchronous, asynchronous and priority updatings are also transitions in the generalized asynchronous dynamics.
Individual bloc-sequential transitions may not correspond to transitions in the generalized asynchronous dynamics; however, equivalent trajectories always exist.
In summary, the reachability properties of the generalized asynchronous dynamics provide an over-approximation of the reachability properties in all other classical updatings.

In addition to the classical updating semantics, the \emph{most permissive} (MP) semantics has recently been proposed to account for
trajectories of multi-valued or continuous refinements which are not captured by the generalized asynchronous dynamics~\citep{pauleve2020most}.
This semantics introduces intermediate states representing uncertainty during the transitions from regular Boolean states: when a component is
in an intermediate state, its target can behave as if it were in either of the classical Boolean state.
In this work, we propose an alternative definition of this semantics in \cref{def:permissive}.
This semantics gives an over-approximation of all classical semantics, including the generalized asynchronous and allows to further recover
additional relevant dynamical trajectories observed in any multi-valued refinements of the Boolean network.
From a computational perspective, while the most permissive semantics increases the cost of explicit simulations due to its large
number of trajectories, it also enables efficient analytical methods for the identification of attractors and reachability properties.

We conclude this section with some additional nomenclature.
For a path or trajectory $P$ in the asynchronous dynamics given by the sequence of states $x^0,\dots,x^l$, we call \emph{direction sequence} of the path $P$ the sequence $i_0,\dots,i_{l-1}$ of the directions of the edges in the path. In other words, the sequences satisfy $x^k_{i_k}\neq x^{k+1}_{i_k}$ for $i=0,\dots,l-1$.
If the direction sequence contains no repetition, we say that $P$ is a \emph{geodesic}.
For convenience, we will call a geodesic in asynchronous dynamics an \emph{asynchronous geodesic}.

A \emph{fixed point} (also called stable state or steady state), is a state $x$ such that $f(x) = x$.
Given a fixed point $x$, we have $\Delta(x,f(x)) = \emptyset$, and this state has no successor in any updating.

An \emph{implicant} of a function is a subspace such that the function is true in all states of the subspace.
An implicant is prime if it is not contained in any larger implicant (i.e. if it has a minimal set of fixed variables).

A \emph{trap space} (also called stable motif), is a subspace $t$ such that for each $x \in t$, $f(x) \in t$. One can think of trap spaces as partial fixed points.
If a state belongs to a trap space, then all its successors in any updating also belong to this trap space. 
We call a trap space minimal if it is not a superset of any other trap space.
Note that the overlap of two trap spaces is also a trap space and that there is a unique minimal trap space containing a given state $x$.

A \emph{trap set} is a subset of the state space that is closed for the dynamics.
An \emph{attractor} is an inclusion-minimal trap set.
It can consist of an isolated state (it is then a fixed point), or of multiple states; in the latter case it is called a \emph{cyclic} or \emph{complex} attractor.
Note that trap sets and attractors may depend on the updating semantics, while fixed points and trap spaces are structural properties of the network itself.
Each trap space is also a trap set and contains at least one attractor for any updating; the number of minimal trap spaces is thus a lower bound for the number of attractors.

\section{Partial orders in asynchronous and permissive trajectories}
\label{sec:partial-orders}

Here we investigate structural conditions for existence of permissive and asynchronous geodesics.
For this, we define \emph{permissive trajectories}, which reproduce the most permissive semantics using classical Boolean states and
subspaces instead of an extended state space based on the addition of transitory states. We will then use implicants associated to
the functions $f$ and their differences with the initial state to identify partial orders enabling permissive geodesics.
The partial orders that satisfy additional constraints correspond to geodesics in the classical asynchronous dynamics.

Given the state $x$ and a subspace $t$, the three following sets of components form a partition of $V$:
\begin{align*}
  \Delta(x,t) &= \{ j \in V ~\mid~ t_j = \neg x_j  \}\text{,} \\
  \Same(x, t) &= \{ j \in V ~\mid~ t_j = x_j \}\text{,}  \\
  \Free(t)    &= \{ j \in V ~\mid~ t_j = \star \}\text{.}
\end{align*}
Observe that the state $x$ is in the subspace $t$ if and only if $\Delta(x,t) = \emptyset$.

\begin{definition}
\label{def:permissive}
A \emph{permissive trajectory} is a succession of states $x^0, x^1, \dots, x^l$ such that for any $k < l$
there is a component $i$ such that $\Delta(x^k, x^{k+1}) = \{i\}$ and the smallest subspace containing all states
$(x^0, \dots, x^k$) contains at least one state $y$ such that $f_i(y) \neq x^k_i$.
By extension, a \emph{permissive geodesic} is a permissive trajectory where each component is used at most once.
\end{definition}

Observe that any classical asynchronous trajectory is a permissive trajectory and that any generalized asynchronous trajectory can also be reproduced
by a permissive trajectory. We can further define a bijection between permissive trajectories and trajectories starting with a pure Boolean
state in the MP semantics.

\begin{proposition}\label{property:2n}
Given any permissive trajectory from $x$ to $y$, there exists a permissive trajectory from $x$ to $y$ of length at most $2n$.
\end{proposition}

This property corresponds to Lemma 1 in the MP supplementary.
Note that we get a bound of $2n$ steps here instead of the $3n$ bound in MP definition as the transitions from
transitory states to regular Boolean states are implicit in the definition of the permissive trajectories.

\begin{proposition}
\label{prop:max-geodesic-trapspace}
  Let $x$ be a state and let $y$ be such that $[x,y]$ is the minimal trap space containing $x$.
  Then all maximal permissive geodesics starting in $x$ end in $y$.
\end{proposition}
\begin{proof}
  Consider a maximal permissive geodesic $P$ from $x$ to a state $z$.
  Since $P$ is maximal, $f_i(w)=x_i$ for all $w \in [x, z]$ and $i \notin \Delta(x,z)$, that is, $[x,z]$ is a trap space, hence it contains $[x,y]$, so $\Delta(x,y) \subseteq \Delta(x,z)$.
  Suppose that $\Delta(x,z) \setminus \Delta(x,y)$ is not empty, and take the first variable $i \in \Delta(x,z)\setminus \Delta(x,y)$ that changes along $P$.
  Then $f_i(w)\neq x_i$ for some $w \in [x, y]$, which contradicts the fact that $[x,y]$ is a trap space.
  Hence $\Delta(x,y)=\Delta(x,z)$, which concludes.
\end{proof}

We are interested in studying reachability from a given an initial condition $x$.
In particular we are interested in determining whether a target state is reachable from $x$ by looking at the implicants defining the network $f$.
To this end, we introduce \emph{implicant maps}, that is, possible choices of implicants for a given target, and give a characterisation of implicant maps that provide paths to the target as either permissive trajectories or asynchronous trajectories.

\begin{definition}
Given a state $x$ and a set of components $J \subseteq V$, the map $\I \colon J \to \{0,1,\star\}^V$ is an \emph{implicant map}
of $J$ for the state $x$ if for each component $i \in J$ and each state $y \in \I(i)$ we have $f_i(y) \neq x_i$.
\end{definition}

Given an implicant map $\I \colon J \to \{0,1,\star\}^V$, we call $\Delta(x, \I(i))$ the set of \emph{direct requirements} of the component $i$ associated to $\I$, and $\Same(x, \I(i)) \setminus \{i\}$ its set of \emph{blockers}.

The set of \emph{strong requirements} $\EDelta(x, \I(i))$ of the component $i$ combines the set of requirements of $i$ with the set of components
blocked by $i$: $\EDelta(x, \I(i)) = \Delta(x, \I(i)) \cup \{ j \neq i ~\mid~ i \in \Same(x, \I(j)) \}$.

Intuitively, we want to establish if an implicant map defines a geodesic from $x$ to $\bar{x}^J$. $\Delta(x, \I(i))$ is the set of components that need to change to enable a change in component $i$. On the other hand, some components can only be updated before a change in component $i$, thus creating some potential ``conflicts'' that forbid some updating orders. The sets $\EDelta$ capture these possible conflicts. To talk about absence of conflicts we introduce the notion of \emph{consistency}.

We need two additional auxiliary constructions.
Given an implicant map $\I\colon J\to \{0,1,\star\}^V$, define the graphs $G(\I,x)$ and $G^+(\I,x)$ with vertex $J$ and edge set
$\{(j,i) ~\mid~ j \in \Delta(x,\I(i))\}$ and $\{(j,i) ~\mid~ j \in \EDelta(x,\I(i))\}$ respectively.

For all $i \in J$, define the sets
\begin{equation*}
  \begin{split}
    \Required(\I, x, i) =  \{j \in J ~\mid~ &\text{there is a path of length}\\ &\text{greater than zero from }j\text{ to }i\text{ in }G(\I,x)\},
  \end{split}
\end{equation*}
\begin{equation*}
  \begin{split}
    \ERequired(\I, x, i) = \{j \in J ~\mid~ &\text{there is a path of length}\\ &\text{greater than zero from }j\text{ to }i\text{ in }G^+(\I,x)\}.
  \end{split}
\end{equation*}
We call $\Required(\I, x, i)$ the set of \emph{full requirements} of $i$
and $\ERequired(\I, x, i)$ the set of \emph{strong full requirements} of $i$.

An implicant map $\I$ is \emph{consistent} if for each $i \in J$ we have $\Required(\I, x, i) \subseteq J \setminus \{i\}$.

An implicant map $\I$ is \emph{strongly consistent} if for each $i \in J$ we have $\ERequired(\I, x, i) \subseteq J \setminus \{i\}$.

The following result establishes that the conditions of consistency and strong consistency exactly characterize the ability of an
implicant map to define a permissive geodesic or an asynchronous geodesic.

\begin{proposition}\label{prop:geodesics}
Given a state $x$ and a set of components $J \subseteq V$, there is a permissive geodesic from $x$ to $\bar{x}^J$ if
and only if there if a consistent implicant map of $J$ for $x$.

Furthermore, there is an asynchronous geodesic from $x$ to $\bar{x}^J$ if and only if there is a strongly consistent
implicant map of $J$ for $x$.
\end{proposition}
\begin{proof}
Take a geodesic $x = x^0, x^1, \dots, x^l = \bar{x}^J$ in the asynchronous dynamics with direction sequence $i_0,\dots,i_{l-1}$.
Consider the map $\I\colon J \to \B^V$ defined by $\I(i_k) = x^k$ for $k=0,\dots,l-1$ (i.e. the map that associates each component
involved in the geodesic with the state in which it changes). Observe that this map is an implicant
map of $J$ for $x$. For each $k=0,\dots,l-1$ we have
\[\Delta(x, \I(i_k)) = \{i_0, \dots, i_{k-1}\}, \ \Same(x, \I(i_k))=J\setminus\{i_0, \dots, i_{k-1}\},\]
\[\EDelta(x, \I(i_k)) = \Delta(x, \I(i_k)) \cup \{ j \neq i_k ~\mid~ i_k \in \Same(x, \I(j)) \} = \{i_0, \dots, i_{k-1}\}.\]
It follows that $\Required(\I, x, i_k) = \ERequired(\I, x, i_k) = \{i_0, \dots, i_{k-1}\}$.
The map $\I$ is thus strongly consistent.

Now we take a permissive geodesic $x = x^0, x^1, \dots, x^l = \bar{x}^J$.
By definition, for all $k=0,\dots,l-1$, there exists $y^k \in [\{x^0,\dots,x^k\}]$ such that $f_{i_k}(y^k) \neq y^k_{i_k}$.
Take the map $\I\colon J \to \B^V$ defined by $\I(i_k) = y^k$ for all $k=0,\dots,l-1$. Observe that $\I$
is an implicant map of $J$ for $x$. In addition, for all $k=0,\dots,l-1$, since $\Delta(x, \I(i_k)) \subseteq \{i_0, \dots, i_{k-1}\}$,
we have $\Required(\I, x, i_k) \subseteq \{i_0, \dots, i_{k-1}\}$. Hence the map $\I$ is consistent.

Consider $\I\colon J \to \{0,1,\star\}^V$ a strongly consistent implicant map of $J$ for $x$ and $G^+(\I,x)$ the associated graph.
Since $\I$ is strongly consistent, we have $i \notin \ERequired(\I, x, i)$ for all $i \in J$, that is, $G^+(\I,x)$ has no cycle.
Hence $G^+(\I,x)$ admits a topological ordering $i_1, \dots, i_l$.
By definition, for each $k \in \{1,\dots,l\}$ the sub-ordering $i_1, \dots, i_{k-1}$ contains all components in $\ERequired(\I,x,i_k)$.
For all $h=1,\dots,k-1$, since $i_h$ precedes $i_k$ in the ordering, we have that $i_k \notin \EDelta(x, \I(i_h))$.
In particular, $i_h$ is not in $\Same(x, \I(i_k))$, and is in either $\Delta(x, \I(i_k))$ or $\Free(\I(i_k))$.
Then for each $k=1,\dots,l$ we have $\bar{x}^{\{i_1, \dots, i_{k-1}\}} \in \I(i_k)$, thus the ordering
defines the asynchronous geodesic $x,\bar{x}^{\{i_1\}},\bar{x}^{\{i_1,i_2\}},\dots,\bar{x}^J$.

The proof for the permissive geodesic case proceeds similarly, with the sets of full requirements replacing the sets of strong full requirements and $G(\I,x)$ replacing $G^+(\I,x)$.

\end{proof}

Let $\I$ and $\I'$ be two different implicant maps for $J$ in $x$. We say that $\I'$ is a \emph{generalization} of $\I$
if for each $i \in J$ we have $\I(i) \subseteq \I'(i)$.
Observe that if $\I$ is (strongly) consistent, then all its generalizations are also (strongly) consistent.
We say that $\I$ is a \emph{prime} implicant map if it has no generalization.
Observe that if $\I$ is a prime implicant map, then for each $i \in J$, $\I(i)$ is a prime implicant of the 
function $f_i$ or of its negation (depending on the value of $x_i$).
In this case, the sets of requirements and blockers, and by extension the (strong) full requirements, associated to each component are minimal.

\begin{lemma}\label{lemma:prime-im}
If $\I$ is a prime implicant map of $J$ for $x$, given $i \in J$:
\begin{itemize}
  \item[(i)] for all $j \in \Delta(x, \I(i))$, the interaction graph of $f$ has an edge from $j$ to $i$;
  \item[(ii)] if $i \in \Same(x, \I(j))$ for some $j\in J$, then the interaction graph of $f$ has an edge from $i$ to $j$.
  In particular, for all $j \in \EDelta(x, \I(i))\setminus \Delta(x, \I(i))$ the interaction graph of $f$ has an edge from $i$ to $j$.
\end{itemize}
\end{lemma}
\begin{proof}
$(i)$ By definition of implicant map, for all $y \in \I(i)$ we have $f_i(y) \neq x_i$.
Consider $j \in \Delta(x, \I(i))$ and suppose that $j \notin \Reg(i)$. Then $\I_j(i) \neq \star$ and $f_i(\bar{y}^j) \neq x_i$ for all $y \in I(i)$.
Then the implicant map $\I'$ defined by $\I'_j(i)=\star$, $\I'_k(i)=\I_k(i)$ for all $k \neq j$ and $\I'(h)=\I(h)$ for all $h \neq i$ is a generalization of $\I$, which contradicts the hypothesis.

$(ii)$ If $j$ is such that $i \in \Same(x, \I(j))$, then $\I_i(j) \neq \star$ and $f_j(y) \neq x_j$ for all $y \in \I(j)$.
If $i$ is not a regulator of $j$, then $f_j(\bar{y}^i) \neq x_j$ for all $y \in \I(j)$ and $\I$ admits a generalization as in the previous point.

\end{proof}

The following proposition is a corollary of the lemma.
Here, given a directed graph $G$, we write $\tilde{G}$ for the undirected graph obtained by ignoring the directions of all edges.

\begin{proposition}
\label{prop:gen-implicant-map}
Consider a Boolean network $(V,f)$ with interaction graph $G$ and $\I \colon J \to \{0,1,\star\}^V$ a prime implicant map for $x$. Then, for all $i,j \in V$:
\begin{itemize}
  \item[(i)] if $j \in \Required(\I, x, i)$ then there is a path of length greater than zero from $j$ to $i$ in $G$;
  \item[(ii)] if $j \in \ERequired(\I, x, i)$ then there is a path of length greater than zero from $j$ in $i$ in $\tilde{G}$;
              if $j \in \ERequired(\I, x, i) \setminus \Required(\I, x, i)$ then there is at least one edge $(h,k)$ in the path such that $(k,h)$ is an edge in $G$.
\end{itemize}
\end{proposition}
\begin{proof}
$(i)$ By~\cref{lemma:prime-im} (i), $G(\I,x)$ is a subgraph of $G$. The conclusion follows from the definition of $\Required(\I, x, i)$.

$(ii)$ By~\cref{lemma:prime-im} (i) and (ii), $\widetilde{G^+(\I,x)}$ is a subgraph of $\tilde{G}$, hence the first part of the statement.
If $j$ is in $\ERequired(\I, x, i)$ but not in $\Required(\I, x, i)$, then at least one of the edges $(h,k)$ in the path satisfies
 $h \in \EDelta(x, \I(k))\setminus\Delta(x, \I(k))$, and we conclude using~\cref{lemma:prime-im} (ii).
\end{proof}

\section{$L$-cuttable Boolean networks}
\label{sec:cuttable-networks}

In the previous section, we identified conditions on the implicant maps associated to a given initial state for the existence of a geodesic
in permissive trajectories or in classical asynchronous trajectories. In presence of a permissive geodesic, we observed that conflicts captured
by the implicant map and the associated auxiliary graph can  prevent the existence of the corresponding asynchronous geodesic. Here we will define
a topological class of networks in which such conflicts do not exist. In this case, all consistent implicant map are also strongly consistent, and 
thus all permissive geodesics exist in the asynchronous dynamics.

In the following, we say that a component of a network is \emph{linear} if it has a single regulator and a single target.
In the next definition we introduce the class of linearly-cuttable networks, that is, networks that admit a set of linear components separating all potential regulatory conflicts.
We will show that, for asynchronous dynamics associated to linearly cuttable Boolean networks, trap spaces provide good approximation of attractors; in addition, we will prove some general reachability properties.

\begin{definition}
\label{def:l-cuttable}
Given a directed graph $G$ on $V$, a \emph{linear cut} of $G$ is a set $L \subseteq V$ of linear components such that
\begin{itemize}
  \item[(i)] every cycle in $G$ contains at least one component of $L$,
  \item[(ii)] every path of length greater than zero in $G$ from a component with multiple targets to a component with multiple regulators contains at least a component of $L$.
\end{itemize}
A linear cut $L$ in \emph{minimal} if there is no linear cut strictly included in $L$.

A Boolean network $M = (V,f)$ is \emph{$L$-cuttable} if $L\subset V$ is a linear cut for its interaction graph $G$.
\end{definition}

We will also need the notion of \emph{canonical states}.
For an $L$-cuttable network $(V,f)$, a state $x \in \B^V$ is \emph{$L$-canonical} if for each $i \in L$ we have $f_i(x) = x_i$.
That is, a state $x$ is $L$-canonical if all components in $L$ are stable in $x$.

Note that if $G$ has a linear cut $L$, then each loop (cycle of length one) is a connected component (since the unique vertex of the loop is necessarily linear). Such a component is called an \emph{isolated loop}. For all the properties we consider in the following, it is easy to see that if $G$ is obtained from $H$ by adding isolated loops, and $H$ satisfies the given properties, then $G$ also satisfies the same properties. Therefore, in all the following, \emph{we assume that $G$ has no loop}.

\begin{remark}
Consider a linear cut $L$ and suppose that there is an edge from $i$ to $j\neq i$ vertices in $L$.
Since $i$ is the unique regulator of $j$, all cycles and all paths in $G$ as in~\cref{def:l-cuttable} (ii) that contain $j$ must also contain $i$.
As a consequence, $L\setminus\{j\}$ is also a linear cut for $G$.
Since we assume that $G$ has no isolated loop, it follows that any minimal linear cut for $G$ is also an independent set of $G$.
In addition, there exists at least one $L$-canonical configuration.
\end{remark}

We now prove properties of implicant maps for networks with linear cuts.

\begin{remark}\label{remark:no-blockers}
  Consider $x$ $L$-canonical and $\I\colon J\to \{0,1,\star\}^V$ a prime implicant map for $x$ and $i \in J\cap L$.
  Since $i$ has only one regulator $j$, if $j\in J$ we must have $\I_j(i)=\bar{x}_j$ and $\I_k(i)=\star$ for all $k \neq j$,
  which gives $\Delta(x, \I(i))=\{j\}$, $\Free(\I(i))=V\setminus\{i\}$ and $\Same(x, \I(i))=\emptyset$.
\end{remark}

\begin{lemma}
Given an $L$-canonical initial state $x$ in an $L$-cuttable network, all consistent implicant maps for $x$ have a strongly consistent generalization.
\end{lemma}
\begin{proof}
Consider a consistent implicant map $\I'$  and take a generalisation $\I$ of $\I'$ that is prime.
Suppose that $\I$ is a consistent but not strongly consistent implicant map for $x$, i.e., there is at least one component
$i$ such that $i \in \ERequired(\I, x, i) \setminus \Required(\I, x, i)$.
By \cref{prop:gen-implicant-map} $(ii)$, $i$ is part of a cycle in $\tilde{G}$,
with at least one edge $(j,k)$ such that $j \in \EDelta(x, \I(k))\setminus\Delta(x, \I(k))$ and $(k,j)\in G$ (at least one edge is associated to a blocker).

If all edges are associated to blockers, the cycle is also a cycle in $G$. By definition of $L$-cuttable network, this cycle contains at least one component of $L$.
As $x$ is canonical, the components of $L$ have no blockers (\cref{remark:no-blockers}) and we have a contradiction.

Thus the cycle contains at least one edge associated to a direct requirement and another edge $(j,k)$ associated to a blocker.
Take the maximal sub-path $\pi$ in the cycle that contains $(j,k)$ and is composed of edges associated to blockers, and call $j'$ and $k'$ the first and last vertex in the path.
Then $G$ contains edges $(j'',j')$ and $(k',k'')$ that are not part of $\pi$, and since the path $\pi$ is associated to blockers, $G$ contains a path from $k'$ to $j'$.
That is, the reverse $\pi'$ of the path $\pi$ is a path in $G$ from a vertex with multiple targets ($k'$) to a vertex with multiple regulators ($j'$).
By definition of $L$-cuttable network, $\pi'$ contains an element of $L$. Since all edges of $\pi'$ are associated to blockers, this is again in contradiction with \cref{remark:no-blockers}.
\end{proof}

By combining the lemma with~\cref{prop:max-geodesic-trapspace} and~\cref{prop:geodesics} we obtain the following.

\begin{corollary}
\label{cor:main}
Let $(V,f)$ be a Boolean network with interaction graph $G$ and $L \subset V$ a linear cut.
All permissive geodesics starting in an $L$-canonical state $x$ exist in the asynchronous dynamics.
In particular:
\begin{itemize}
  \item[(i)] $[x, y]$ is the minimal trap space containing $x$ if and only if there exists a maximal geodesic from $x$ to $y$.
  \item[(ii)] for all subsets of components $J \subseteq \Delta(x, f(x))$ there exists a path from $x$ to $\bar{x}^J$ (all the successors in the generalized asynchronous dynamics are reachable from $x$).
  \item[(iii)] The smallest subspace containing the states that are reachable from $x$ is a trap space.
  \item[(iv)] The smallest subspace containing an attractor is a trap space.
  \item[(v)] If $x$ belongs to an attractor $A$, there is a geodesic from $x$ to $\bar{x}^{\Delta(A)}$, and $\bar{x}^{\Delta(A)}$ is $L$-canonical.
  \item[(vi)] If $y$ is the last vertex of a geodesic starting from $x$ and $f_i(z)\neq z_i$ for some $z\in [x,y]$ and $i \notin \Delta(x,y)$, then there is a geodesic from $x$ to $\bar{y}^i$.
\end{itemize}
\end{corollary}

The conclusions of the corollary do not hold for states that are not canonical: for instance, in the asynchronous dynamics of the Boolean network with two variables defined by $f(x_1,x_2)= (x_2,x_1)$ there are no paths from the non-canonical state $01$ to $10$, while there are transitions to $00$ and $11$.
This example also shows that point $(i)$ of~\cref{def:l-cuttable} in cannot be relaxed.

\subsection{Reachability of trap spaces from canonical states}

\begin{lemma}\label{lem:subgeo}
Let $(V,f)$ be a Boolean network and $P$ a geodesic from $x$ to $y$ with direction sequence $w$.
Let $i\in \Delta(x,y)$ and suppose that $G$ has no edge from $i$ to a vertex that appears after $i$ in $w$.
Then there exists a geodesic from $x$ to $\bar{y}^i$ whose direction sequence is obtained from $w$ by deleting $i$.
\end{lemma}

\begin{theorem}\label{thm:reachability}
Let $(V,f)$ be a Boolean network with interaction graph $G$ and $L \subset V$ a linear cut.
Let $x$ be an $L$-canonical configuration and $[x,y]$ be the minimal trap space containing $x$.
For every trap space $t \subseteq [x,y]$ there is a path in the asynchronous dynamics from $x$ to $t$ of length at most $2n$.
\end{theorem}
\begin{proof}
Define $J=\Delta(x, t)\subseteq \Delta(x, y)$.
By~\cref{cor:main} $(i)$, there is a geodesic from $x$ to $y$.
Take $z \in [x,y]$ such that $J \subseteq \Delta(x, z)$, there is a geodesic $P$ from $x$ to $z$ and the cardinality of $K = \Delta(z, t)$ is minimal.

Since $K\cap J=\emptyset$, for any $i\in K$ we have $z_i \neq x_i$ and thus $i$ appears in the direction sequence of $P$. Let $i_0,\dots,i_{l-1}$ be an enumeration of $K$ as in the direction sequence of $P$.

We first prove the following property.
\begin{quote}
(I) {\em There is no $0\leq p\leq  q\leq l$ such that $G$ has an edge from $i_q$ to $i_p$.}

Suppose, for a contradiction, that there is $1\leq p\leq q\leq l$ such that $G$ has an edge from $i_q$ to $i_p$. Since $G$ has no loop we have $p<q$. Suppose first that $i_p$ has only one regulator.
Since $i_q$ is in $K$ and not in $J$, we have $x_{i_q}=t_{i_q}$ and since $t$  a trap space and $i_q$ is the unique regulator of $i_p$, we derive $f_{i_p}(x)=x_{i_p}$.
Then $i_q$ appears before $i_p$ in the direction sequence of $P$, a contradiction.
So $i_p$ has at least two regulators. Since $G$ has a linear cut, $i_p$ is the unique target of $i_q$, and we deduce from \cref{lem:subgeo} that there is a geodesic from $x$ to $\bar{z}^{i_q}$. Since $i_q$ is in $K$, this contradicts the minimality of $K$.
\end{quote}

Let us prove that there is a geodesic $z=z^0,z^1,\dots,z^l=\bar{z}^K$ from $z$ to $\bar{z}^K$ with direction sequence $i_0,\dots,i_{l-1}$. We have to prove that $f_{i_k}(z^k)\neq z^k_{i_k}$ for $0\leq k<l$.
Since $t$ is a trap space, $f_j(y)=y_j \neq z_j$ for all $j \in K$, therefore it is sufficient to show that $z^{k}_j=t_j$ for any regulator $j$ of $i_k$.

We proceed by induction on~$k$. Let $j$ be a regulator of $i_0$. By (I) we have $j\not\in K$, so $z^0_j=z_j=t_j$.
Let $0<k<l$ and let $j$ be a regulator of $i_k$. If $j\not\in K$, then $z^k_j=z_j=t_j$ by definition of $K$. Otherwise, by (I) we have $j\in\{i_0,\dots,i_{k-1}\}$ thus $z_j\neq z^k_j$, and we deduce that $z^k_j=t_j$.
\end{proof}

\begin{example}\label{ex:not-all-reach-attractors}
  The theorem does not hold if the initial state is not $L$-canonical.
  The Boolean network $f(x_1,x_2,x_3,x_4,x_5) = (x_3, x_4\wedge x_5, x_1, x_1, x_2)$ is $L$-cuttable with $L = \{3,4,5\}$.
  The fixed points of $f$ are $00000$, $10110$ and $11111$.

  Consider the state $x=11011$, which is not $L$-canonical ($f(x)_1 \neq x_1$).
  The fixed point $11111$ is a direct successor for $x$ in the asynchronous dynamics of $f$.
  In addition, $00000$ is reachable from $x$ in the asynchronous dynamics of $f$ via the path
  $11011 \to 01011 \to 01001 \to 00001 \to 00000$.
  As a consequence, the minimal trap space containing $x$ is the full space $\B^5$.
  Observe that there is no path from $11011$ to the fixed point $10110$.
\end{example}

\subsection{Minimal trap spaces are good approximations for attractors}

In this section we prove that, in asynchronous dynamics of linearly-cuttable networks, attractors and minimal trap spaces are in one-to-one correspondence.

\begin{theorem}\label{thm:attractor}
  Suppose that $(V,f)$ is $L$-cuttable and $A$ is an attractor for the asynchronous dynamics of $f$.
  Then $[A]$ is a trap space and, for every $x\in [A]$, there is a geodesic from $x$ to $A$.
\end{theorem}

Given two configurations $x,y$, we set $[x,y[=[x,y]\setminus\{y\}$.

\begin{lemma}\label{lem:geo_xy}
  Suppose that $(V,f)$ is $L$-cuttable and $A$ is an attractor for the asynchronous dynamics of $f$.
  Let $x\in [A]$ and $y\in A$, and suppose that $y$ is $L$-canonical. Let $I$ be the set of $i\in L$ with $f_i(x)\neq x_i=y_i$. Suppose that there is no $L$-canonical configuration in $[\bar{x}^I,y[\cap A$. Then there is a geodesic from $x$ to $y$.
\end{lemma}
\begin{proof}
We proceed by induction on  the Hamming distance $d(x,y) = \norm{\Delta(x,y)}$. If $d(x,y)=0$ there is nothing to prove, so suppose that $d(x,y)>0$. We need the following.

\begin{quote}
(1) {\em There is no $i\in\Delta(x,y)\setminus L$ such that $\bar{y}^i\in A$.}

Suppose, for a contradiction, that $z=\bar{y}^i$ is in $A$ for some $i\in\Delta(x,y)\setminus L$. Let $J$ be the targets $j$ of $i$ such that $j\in L$ and $f_j(z)\neq z_j$. Since $y$ is $L$-canonical and $L$ is an independent set, there is a geodesic from $z$ to $\bar{z}^J$, which is $L$-canonical. Suppose, for a contradiction, that $\bar{z}^J\not\in [\bar{x}^I,y]$. Then there is a component $j$ such that $\bar{z}^J_j\neq \bar{x}^I_j=y_j$. Since $x_I=y_I$ we have $j\not\in I$, thus $\bar{z}^J_j\neq x_j=y_j$. Since $x_i\neq y_i$ we have $j\neq i$, thus $\bar{x}^J_j\neq x_j=y_j$. We deduce that $j\in J$. Since $x_i=z_i\neq y_i$ and since $y$ is $L$-canonical, we have $f_j(x)=f_j(z)\neq f_j(y)=y_j=x_j$ thus $j\in I$, a contradiction. This proves that $\bar{z}^J\in [\bar{x}^I,y]$, and since $z_i\neq y_i$ we have $z\in [\bar{x}^I,y[$. Since $\bar{z}^J$ is $L$-canonical and reachable from $y$, we have $\bar{z}^J\in A$ and we obtain a contradiction.
\end{quote}

\begin{quote}
(2) {\em $f_i(x)\neq x_i$ for some $i\in\Delta(x,y)$.}

Suppose not, that is, $f_i(x)=x_i$ for all $i\in\Delta(x,y)$. Since $y$ is $L$-canonical, by~\cref{cor:main} $(v)$, there is a geodesic from $y$ to $y'=\bar{y}^{\Delta(A)}$, and a geodesic $P$ from $y'$ to $y$. Let $i$ be the first component of the direction sequence of $P$ with $x_i\neq y_i$. Since $x,y\in [A]$, we have $\Delta(x,y)\subseteq \Delta(A)$, thus this component $i$ exists. Let $z$ be the configuration of $P$ with $f_i(z)\neq z_i$.

Let us prove that $i$ has at least two regulators. Suppose not. Since $i\in \Delta(A)$, $i$ has only one regulator, and its regulator $j$ is in $\Delta(A)$. If $i\in L$ then $f_i(y)=y_i$ since $y$ is $L$-canonical, and if $i\not\in L$, then, since $x_i\neq y_i$, we have $f_i(y)=y_i$ by (1). Since $i,j\in\Delta(A)$, we obtain $f_i(y')=y'_i$. Since $f_i(z)\neq z_i=y'_i$, we have $z_j\neq y'_j$ and thus $j$ appears before $i$ in the direction sequence of $P$. By the choice of $i$, we have $x_j=y_j$ and thus $f_i(x)=f_i(y)=y_i\neq x_i$, which contradicts our hypothesis. This proves that $i$ has at least two regulators.

Let $P'$ be the path from $z'=\bar{z}^i$ to $y$ contained in $P$. Let $J$ be the set of regulators $j$ of $i$ such that $x_j\neq y_j$. We have $i\not\in J\subseteq\Delta(x,y)\subseteq\Delta(A)$. Hence, by the choice of $i$, $J\cap \Delta(y',z')=\emptyset$, and since $\Delta(y',z'),\Delta(z',y)$ is a partition of $\Delta(A)$, we have $J\subseteq \Delta(z',y)$. Hence $\bar{y}^J\in [z',y]$. By the definition of $J$ and our hypothesis, we have $f_i(\bar{y}^J)=f_i(x)=x_i\neq y_i=\bar{y}^J_i$. Since $z'_i=y_i$, we deduce from~\cref{cor:main} $(vi)$ that there is a geodesic from $z'$ to $\bar{y}^i$, and since $\bar{y}^i$ is reachable from $y$, we have $\bar{y}^i\in A$. Since $i$ has at least two regulators, this contradicts (1).
\end{quote}

\medskip
By (2) there is a component $i$ with $x_i\neq f_i(x)=y_i$. Then there is a transition from $x$ to $z=\bar{x}^i$.  Let $J$ be the set of $j\in L$ with $f_j(z)\neq z_j=y_j$ (we have $i\not\in J$ since otherwise $i$ has a negative loop). Let us prove that $\bar{z}^J\in [\bar{x}^I,y]$.

Take a component $j$ such that $\bar{x}^I_j=y_j$. We have to show that $\bar{z}^J_j=\bar{x}^I_j=y_j$. We have $j \notin I$ by definition of $I$, hence $x_j=y_j$. Since, by choice of $i$, $x_i \neq y_i$, we have $j \neq i$, so $z_j=\bar{x}^i_j=x_j=y_j$. Suppose that $j$ is in $J$, that is, $j \in L$ and $f_j(z) \neq z_j$. Since $j$ is not in $I$, we have $f_j(x)=x_j$, and $i$ is therefore the unique regulator of $j$. Since $z_i=y_i$, we have $f_j(y)=f_j(z)\neq y_j$, but then $y$ is not $L$-canonical, a contradiction. Hence $j$ is not in $J$ and $\bar{z}^J_j=\bar{x}^I_j=y_j$ as wanted.

This proves that $\bar{z}^J\in [\bar{x}^I,y]$ and thus $[\bar{z}^J,y]\subseteq [\bar{x}^I,y]$. Hence, by hypothesis, there is no $L$-canonical configuration in $[\bar{z}^J,y[\cap A$. Since $d(z,y)<d(x,y)$, by induction, there is a geodesic from $z$ to $y$ and thus it has also a geodesic from $x$ to $y$.
\end{proof}

\cref{thm:attractor} follows from \cref{cor:main} $(iv)$ and the next lemma.

\begin{lemma}\label{lem:geo_xyz}
  Suppose that $(V,f)$ is $L$-cuttable and $A$ is an attractor for the asynchronous dynamics of $f$.
  Let $x\in [A]$ and $y\in A$, and suppose that $y$ is $L$-canonical. Let $I$ be the set of $i\in L$ with $f_i(x)\neq x_i=y_i$.
  Then there is a geodesic from $x$ to some $L$-canonical configuration $a \in [\bar{x}^I,y]\cap A$.
\end{lemma}

\begin{proof}
We proceed by induction on $d(\bar{x}^I,y)$. Since $x_I=y_I$, we have $x\in [\bar{x}^I,y]$, so if $d(\bar{x}^I,y)=0$ then $x=y$ and there is nothing to prove. So suppose that $d(\bar{x}^I,y)>0$. If there there is no $L$-canonical configuration in $[\bar{x}^I,y[\cap A$, then, by \cref{lem:geo_xy}, there is a geodesic from $x$ to $y$, so the lemma holds with $a=y$. So suppose that there is an $L$-canonical configuration $y'\in [\bar{x}^I,y[\cap A$. Let $I'$ be the set of $i\in L$ with $f_i(x)\neq x_i=y'_i$.

\medskip
We have $[\bar{x}^{I'},y']\subseteq [\bar{x}^I,y]$. Indeed, since $y'\in [\bar{x}^I,y]$ it is sufficient to prove that $\bar{x}^{I'}\in [\bar{x}^I,y]$.
That is, given $i$ such that $\bar{x}^{I}_i=y_i$, we have to show that $\bar{x}^{I'}_i=\bar{x}^I_i$.
Since, by definition of $I$, we have $i \notin I$, we just need to show that $i$ is not in $I'$.
Since $y'$ is in $[\bar{x}^I,y]$, we have $y'_i=y_i$; as a consequence, $i \in I'$ would imply $i \in I$, a contradiction.

\medskip
We have $\Delta(y,y')\setminus L\neq\emptyset$. Indeed, let $i\in\Delta(y,y')$. If $i\not\in L$ we are done. So suppose that $i\in L$ and let $j$ be one of its regulators. Since $y',y$ are $L$-canonical, $f_i(y')=y'_i\neq y_i=f_i(y)$ thus $y'_j\neq y_j$. Since $L$ is a minimal linear cut, $L$ is an independent set thus $j\not\in L$ so $j\in\Delta(y,y')\setminus L$. 

\medskip
So let $i\in\Delta(y,y')\setminus L$. Since $I',I\subseteq L$ and $y'\in [\bar{x}^I,y]$ we have $y_i\neq y'_i=\bar{x}^I_i=\bar{x}^{I'}_i$, thus $i\in\Delta(\bar{x}^I,y)\setminus\Delta(\bar{x}^{I'},y')$. Since $[\bar{x}^{I'},y']\subseteq [\bar{x}^I,y]$ we have $\Delta(\bar{x}^{I'},y')\subseteq\Delta(\bar{x}^I,y)$ and we deduce that $d(\bar{x}^{I'},y')<d(\bar{x}^I,y)$.

\medskip
Consequently, by induction, there is an $L$-canonical configuration $a\in [\bar{x}^{I'},y']\cap A\subseteq [\bar{x}^I,y]\cap A$ such that there is a geodesic from $x$ to $a$. This completes the induction.
\end{proof}

\begin{remark}
  \cref{thm:attractor} shows that every linearly-cuttable network has the property that
  each minimal trap space contains only one attractor.
  While attractors of most permissive semantics coincide with minimal trap spaces~\citep{pauleve2020most},
  this is not always true for linearly-cuttable networks,
  as can be seen for instance by taking Boolean networks with interaction graph
  consisting of a negative cycle \citep[see][for a full
  characterisation of the dynamics associated to isolated circuits]{remy2003description}.
\end{remark}

\section{Cuttable extended semantics}
\label{sec:semantics}

\begin{figure}
  \begin{center}
    \includegraphics[width=\figwidth]{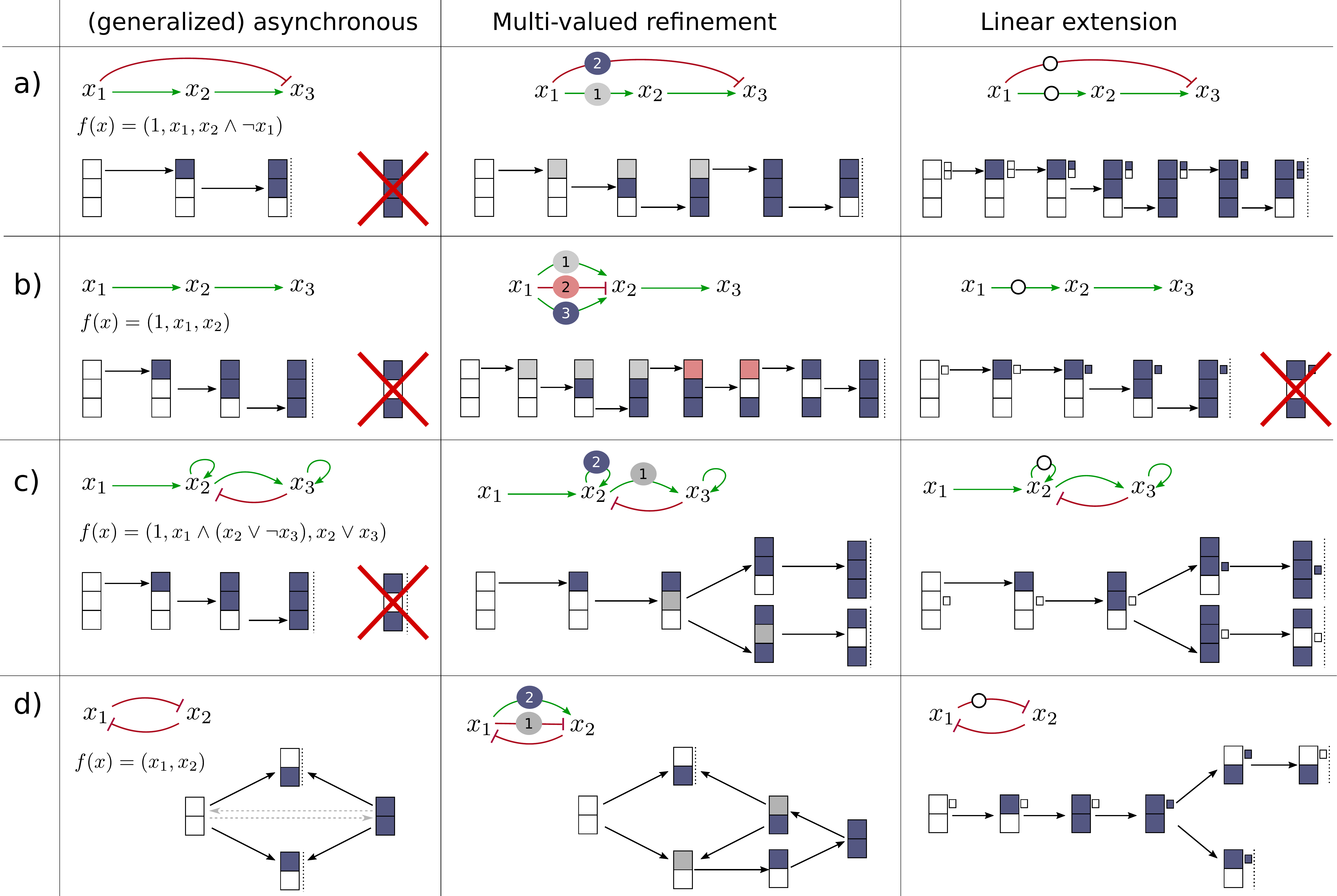}
  \end{center}
  \caption{\label{fig:trajs}Reachability properties in Boolean, refined and extended networks.\\
Each row shows a Boolean network with its asynchronous dynamics (left), one of its multi-valued refinements (center) and linear extensions (right).
White circles in the extended network denote intermediate linear variables, whereas numbered coloured circles in the refined network denote regulatory thresholds.
Selected dynamical trajectories are depicted below each interaction graph. Groups of color-coded squares represent the states of all variables:
white for level 0, blue for level 1 (or max), gray and red denote intermediate levels in refinements. Fixed points are marked with a dotted line on the right.
The values of intermediate linear variables are represented with smaller squares on the right side of their regulators.
\textbf{a)} An inconsistent feedforward loop: the first component has opposite (direct and undirect) effects on the last one.
    This competition can be relaxed by associating a higher threshold (center) or adding an intermediate component (left) to the direct interaction.
\textbf{b)} A chain propagating an activation. In the most permissive semantics and some non-monotonic refinements (center), intermediate components can be disabled
    after propagating the signal. This behaviour can often be considered as an artefact and can not be reproduced in linear extensions.
\textbf{c)} A chain with stabilizing feedback loops. This is an extension of the previous example where feedback loops are added to stabilize the unexpected $(1,0,1)$ state.
    This state is still unreachable in the Boolean network, however it can now be reached in monotonic (single threshold) refinements and in linear extensions.
\textbf{d)} A positive circuit showing that the reachability of the generalized asynchronous (where transitions can involve multiple components) can be reproduced in 
    linear extensions, however it may not be faithfully reproduced in multi-valued refinements.
}
\end{figure}

Given a Boolean network, we obtain an \emph{extended network} by replacing a subset of the interactions with linear components.
We show that the trap spaces of the original network are also trap spaces of its extensions, which provide an over-approximation of the original asynchronous dynamics.
We will focus on \emph{cuttable extended networks} in which the additional linear components form a linear cut of the extended network.
Cuttable extensions allow to define an execution semantics that takes advantage of the properties of cuttable networks for any Boolean network.

Biological Boolean networks are abstract models often used in absence of quantitative knowledge on precise concentrations and kinetic
parameters. The non-determinism of the classical asynchronous semantics accounts for this lack of knowledge by enabling alternative
trajectories corresponding to quantitative differences in initial conditions and kinetic parameters. However, it assumes that a change
of the state of a component is reflected on all its targets at the same time. The introduction of intermediate linear components lets
us eliminate this assumption. The alternative trajectories obtained in the asynchronous dynamics of an extended network then cover
plausible behaviours that may be missing in the asynchronous dynamics of the original network.

\begin{definition}
  Let $M = (V,f)$ be a Boolean network with edges $E$ and $L \subseteq E \subseteq V^2$ a subset of its interactions.
  Consider the Boolean function $\extension(f,L)\colon \B^{V \cup L} \to \B^{V \cup L}$ defined as follows. For each $i \in V \cup L$
  \begin{align*}
  \extension(f,L)_i(y) & = \begin{cases}
    f_i( \pi^i(y) ) & \text{ if } i \in V, \\
    y_j & \text{ if } i = (j,k) \in L,
    \end{cases}
  \end{align*}
  where $\pi^i\colon \B^{V \cup L} \to \B^V$ is defined for all $j\in V$ as:
  \begin{align*}
   \pi^i(y)_j & = \begin{cases}
    y_{(j,i)} & \text{ if } (j,i) \in L, \\
    y_j     & \text{ otherwise.}
  \end{cases}
  \end{align*}
  
  We call the Boolean network $(V \cup L, \extension(f,L))$ an \emph{extended network} and the \emph{$L$-extension} of $f$.
\end{definition}

For an extended network $(V \cup L, \extension(f,L))$, we call $V$ the set of \emph{core variables} and $L$ the set of \emph{extender variables}.
We say that an $L$-extension \emph{cuttable} if it is $L$-cuttable.
We call the $E$-extension of $f$ its \emph{full extension}. By construction, the $E$-extension is cuttable.
We will need the following additional notations. We write $\pi\colon \B^{V \cup L} \to \B^V$ for the projection onto $\B^V$,
and define the map $\E \colon \B^V \to \B^{V \cup L}$ that ``copies'' each regulator, once for each of its target variable:
\begin{equation*}
  \E_k(x) = \begin{cases}
    x_j & \text{ if } k = (j,i) \in L, \\
    x_k & \text{ otherwise.}
  \end{cases}
\end{equation*}

Note that $x = \pi( \E(x) ) = \pi^i( \E(x) )$ for any $i \in V$, and that if $L$ contains no interaction with target $i$, then $\pi^i = \pi$.
We call the states $y \in V \cup L$ that satisfy $\E(\pi(y))=y$ (that is, states for which the extender variables mirror their regulators) \emph{canonical} states.
Note that, by construction, all canonical states of an $L$-extended network are $L$-canonical.

Aside from the partition of their components into core and extender variables, extended networks are regular networks and the notations
introduced above, such as $T(i)$ and $R(i)$, apply as usual. Depending on the context, extender variables will be referred to as regular
variables (e.g. $i \in (V \cup L)$) or as a pair of core variables (e.g. $(i,j) \in V^2$).

\begin{definition}
Let $M = (V,f)$ be a Boolean network, $x$ and $y$ two states of $\B^V$, and $L$ a subset of its interactions.
We say that $y$ is \emph{$L$-reachable} from $x$ if there is a trajectory from $\E(x)$ to $\E(y)$ in the asynchronous
dynamics of the $L$-extension of $M$.
\end{definition}

This definition of $L$-reachability allows us to study reachability in any Boolean network using canonical initial states in an extended network.
Note that the set of states that are reachable from a non-canonical state can differ significantly from the set of states that are reachable
from  the canonical state that projects to the same core variables. For instance, consider a Boolean network such that all components have at
least one regulator, and take the full extension. Then all canonical states are reachable from any state in which all extender variables differ
from their regulators.

It is worth observing that the elimination of the extender components from the extended network using the method described in
\citet{Naldi2011reduction} allows to recover the original network. The asynchronous dynamics of an extended network is thus
an over-approximation of the original asynchronous dynamics. As consequence, If $y$ is $L$-reachable from $x$,
then it is also $K$-reachable for any $K \supset L$.

In the following we compare in more detail the reachability properties of the  original network and its cuttable extensions and relate
the trap spaces of a Boolean network $(V,f)$ to the trap spaces of its $L$-extension.

Observe that the image under $\E$ of a subspace $[x,y] \subseteq \B^V$ is the subspace $\E([x,y])=[\E(x), \E(y)]$
with $\Delta(\E(x),\E(y)) = \Delta(x,y) \cup I'$ where $I'$ is the subset of extender variables $\{(j,i)\in L$ 
such that $j\in \Delta(x,y)\}$.
By extending the terminology from states to subspaces, we call subspaces of this form \emph{canonical}.

\begin{proposition}
\label{prop:tps}
  Consider a Boolean network $(V,f)$ and its $L$-extension $(V \cup L, f^L)$.
  \begin{itemize}
    \item[(i)] If $[x,y]$ is a trap space for $f$, then $\E([x,y])$ is a canonical trap space for $f^L$.
               If $[x,y]$ is the minimal trap space containing $x$, then $\E([x,y])$ is the minimal trap space containing $\E(x)$.
    \item[(ii)] If $[x',y']$ is a trap space for $f^L$, then $[\pi(x'), \pi(y')]$ is a trap space for $f$ and $\Delta(\pi(x'), \pi(y')) = \Delta(x',y') \cap V$.
    If $[x',y']$ is the minimal trap space containing $x'$, then $[\pi(x'), \pi(y')]$ is the minimal trap space containing $\pi(x')$.
  \end{itemize}
\end{proposition}
\begin{proof}
(I) The fact that subspaces $\E([x,y])$ and $[\pi(x'), \pi(y')]$ are trap spaces is a direct consequence of the definitions of $f^L$, $\E$ and $\pi$.

(II) Suppose that $[x,y]$ is minimal, and consider a trap space $[w',z']$ contained in $[\E(x),\E(y)]$, that is, such that $\Delta(w',z') \subseteq \Delta(\E(x), \E(y))$.
We have to show that $[w',z']=[\E(x),\E(y)]$.
By point (I), $[\pi(w'), \pi(z')]$ is a trap space contained in $[x,y]$, hence it coincides with $[x,y]$.
As a consequence, $\Delta(\pi(w'),\pi(z'))=\Delta(w',z')\cap V=\Delta(x,y)$.
Consider $(j,i) \in \Delta(\E(x), \E(y)) \cap L$, then $j\in \Delta(x,y)=\Delta(\pi(w'),\pi(z'))$ by definition.
Since $[w',z']$ is a trap space, by definition of $f^L$ we have $(j,i) \in \Delta(w', z')$.
Hence $\Delta(w', z') = \Delta(\E(x), \E(y))$, which concludes.

(III) Suppose now that $[x',y']$ is a minimal trap space for $f^L$; we show that $[\pi(x'), \pi(y')]$ is minimal.
Consider a trap space $[z, t]$ contained in $[\pi(x'), \pi(y')]$. Then, by point (I), $\E([z, t])$ is a trap space contained in $[x', y']$, hence coincides with $[x', y']$.
As a consequence, their projections $\pi(\E([z, t]))=[z,t]$ and $[\pi(x'), \pi(y')]$ are equal.
\end{proof}

The proposition states that all trap spaces in extended networks project to trap spaces for the original network,
and any trap space in the original network gives at least one trap space in any extension.
In addition, if $y$ is a canonical state in an extended network, that is $y=\E(x)$ for some $x$,
then the minimal trap space containing $y$ is the canonical extension of the minimal trap space containing $x$.

Clearly a Boolean network and its extensions do not necessarily have the same number of trap spaces. Multiple trap spaces in an
extension can project to the same trap space in the original network. Take for instance the Boolean network $f(x_1)=x_1$ and its
extension $f^L(x_1,x_2)=(x_2, x_1)$ with $L=(1,1)$. The trap spaces $00$ and $0\star$ for $f^L$ project on the same trap space (the fixed point $0$).
On the other hand, the mapping between trap spaces described in the proposition defines a one-to-one correspondence between minimal
trap spaces of a Boolean network and any of its extensions.

\begin{corollary}\label{cor:one-to-one-min-trap}
  There is a one-to-one correspondence between the minimal minimal trap spaces of a Boolean network and the minimal trap spaces of any of its extensions.
\end{corollary}

\begin{remark}
 Minimal trap spaces in extended networks are always canonical. Every trap space $T$ in an extended network contains the canonical trap space $\E(\pi(T))$.
\end{remark}

We now focus our study on cuttable extensions. As stated above, the full extension is always cuttable, but other cuttable extensions often exist in practice.
Following the definition of cuttable networks, these more conservative cuttable extensions can be obtained by extending only interactions $(i,j)$ such 
that $\norm{T(i)} > 1$ and $\norm{\Reg(j)} > 1$ as well as one interaction for each cycle which remains unextended. The following properties build on the
previous results obtained on cuttable networks and can be applied to any cuttable extension.

\begin{proposition}\label{proposition:cuttable-semantics}
Let $M$ be a Boolean network and $L$ a subset of its interactions defining a cuttable extension.
\begin{itemize}
  \item[(i)]   If there is a trajectory from $x$ to $y$ in the generalized asynchronous dynamics of $M$, then $y$ is $L$-reachable from $x$.
  \item[(ii)]  Given a state $x$ and $t$ the minimal trap space containing $x$, all trap spaces contained in $t$ are $L$-reachable from $x$.
  \item[(iii)] There is a one-to-one correspondence between the minimal trap spaces of $M$ and the attractors in the asynchronous dynamics of its $L$-extension.
\end{itemize}
\end{proposition}
\begin{proof}
  $(i)$ It is sufficient to show that, if $\bar{x}^J$ is a successor of $x$ in the generalized asynchronous dynamics of $M$,
  then $\bar{x}^J$ is $L$-reachable from $x$.
  By definition of extended network we have, for all $i\in J$, $\extension(f, L)_i(\E(x))=f_i(x)\neq x_i=\E_i(x)$,
  and $\overline{\E(x)}^J$ is a successor of $\E(x)$ in the generalized asynchronous dynamics of the extended network.
  By \cref{cor:main} (ii), $\overline{\E(x)}^J$ is reachable from $\E(x)$ in the asynchronous dynamics of the extended network.
  Since $\overline{\E(x)}^J$ and $\E(\bar{x}^J)$ coincide on the core variables and $\E(\bar{x}^J)$ is canonical, $\E(\bar{x}^J)$
  can be reached from $\overline{\E(x)}^J$. Combining the two paths we have that $\E(\bar{x}^J)$ is reachable from $\E(x)$.

  $(ii)$ Consider a trap space $t'$ contained in $t$.
  By~\cref{prop:tps}, $\E(t')$ is a trap space contained in $\E(t)$, and $\E(t)$ is the minimal trap space cointaining $\E(x)$.
  \cref{thm:reachability} then gives that $\E(t')$ is reachable from $\E(x)$ in the extended network, that is,
  there exists $y \in \E(t')$ such that there is a path from $\E(x)$ to $y$ in the asynchronous dynamics of the extended network.
  In addition, we can assume that $y$ is canonical, that is, $\E(\pi(y))=y$.
  Then $\pi(y)$ is in $t'$ is $L$-reachable from from $x$.

  $(iii)$ Consequence of~\cref{thm:attractor} and~\cref{cor:one-to-one-min-trap}.
\end{proof}

\begin{figure}
  \begin{center}
    \includegraphics[width=\figwidth]{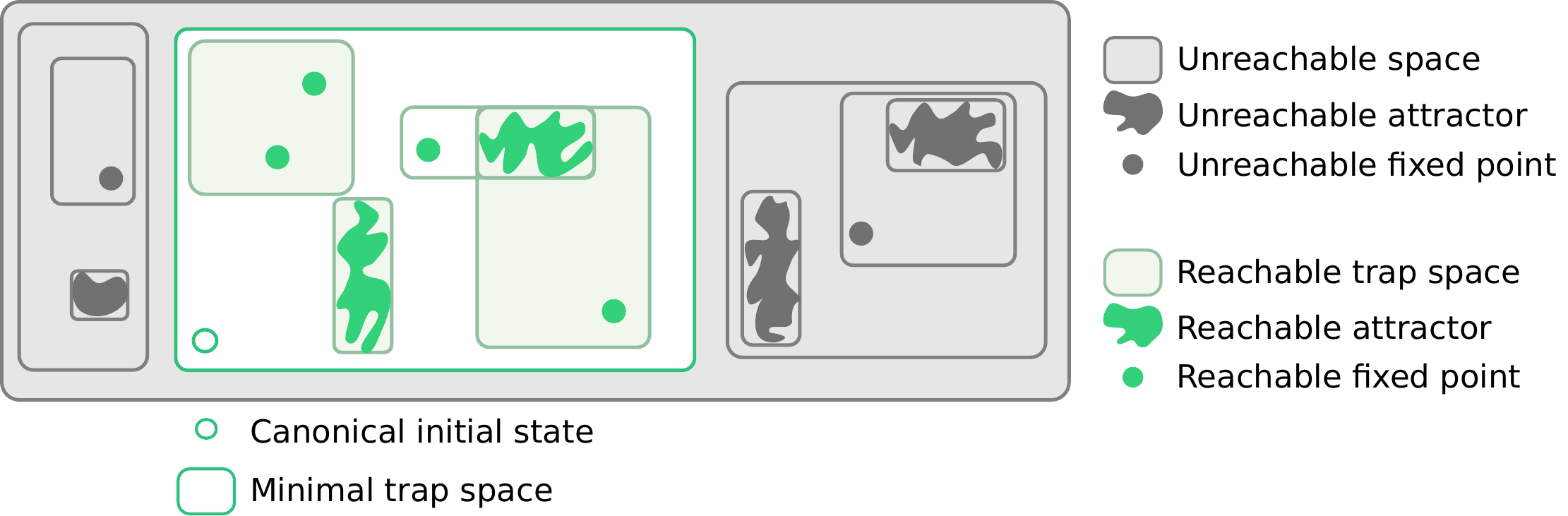}
  \end{center}
  \caption{\label{fig:trapspaces}Summary of the reachability of trap spaces and attractors.
  Given an initial state, all states, and in particular all attractors, that are not contained
  in the minimal trap space containing the initial state are not reachable
  in any updating semantic. For $L$-cuttable networks and $L$-canonical initial states,
  all trap spaces and attractors included in the minimal trap space are reachable.
  }
\end{figure}

\subsection{Relation to single threshold refinements}

Multi-valued networks are commonly used to refine the behaviour of some components of a Boolean network.
They can account for some semi-quantitative knowledge, for instance by tracking different amounts of a component that are
required to affect its different targets, or by encoding the existence of some specific condition leading to a higher production
or a higher activity level for some target. To account for all these effects, multi-valued refinements can take many forms
and involve complex modifications to the logical rules \citep{chaouiya2003qualitative}. Here we introduce \emph{single threshold} networks,
a subset of multi-valued networks that adds different thresholds to the interactions but retains
the same logical rules as the Boolean network. Such refinements are solely defined by a Boolean network and
a mapping associating a single multi-valued threshold to each interaction of the network.

We start by setting some notation and definitions.
Given a Boolean network $M = (V, f)$ with $V = \{1, \dots ,n\}$, we call any $\T \colon V^2 \to \N^*$ a \emph{threshold map} for $M$.
For each $i \in V$, we then define the value $m^i$ and the mapping $\mvtobool^i \colon \N^V \to \B^V$ such that:
\begin{align*}
  m^i              & = \max(\{1\} \cup \{ \T(i,j) ~\mid~ j \in \Target(i) \}), \\
  \mvtobool^i(x)_j & = \ind( x_j \geq \T(j,i) ) \text{ for each } j \in V.
\end{align*}

We call $\M = \prod_{i \in V} [0,m^i]$ the \emph{multi-valued space} of $(M,\T)$.
For each component $i$, we denote by $\mathrm{e}^i$ the element of $\M$ with component $i$ equal to $1$ and all other components equal to $0$.
In addition, we define the mapping $\booltostr \colon \B^V \to \M$ such that for each component $i \in V$, $\booltostr(x)_i = m^i \cdot x_i$.

\begin{definition}
\label{def:str}
Given Boolean network $M = (V,f)$ and a threshold map $\T$ for $M$, the function
\begin{align*}
  \refinement(f, \T) & \colon \M \to \M \\
  \refinement(f, \T)_i&=\booltostr_i \circ f \circ \mvtobool^i \ \text{ for all } i \in V
\end{align*}
is called the \emph{$\T$-refinement} of $M$.
The multi-valued network $\mathcal{M} = (\M, \refinement(f, \T))$ is a \emph{single threshold refinement} of $M$.
\end{definition}

As is customary for multi-valued networks we consider dynamics that allow for asynchronous stepwise transitions that point in the direction defined by the multi-valued function.
That is, we define the asynchronous dynamics of $\mathcal{M}$ as the graph with vertex set $\M$ and edge set
$\{(x, x+\varepsilon \mathrm{e}^i)\ \mid \ x \in \M, \ i \in \Delta(x, \refinement(f, \T)(x)), \ \varepsilon=\mathrm{sign}(\refinement(f, \T)_i(x)-x_i)\}$.

\begin{proposition}\label{prop:boolinstr}
Let $M$ be a Boolean network and $\T$ a threshold map for $M$.
If there exists a transition $x \to \bar{x}^i$ in the asynchronous dynamics of $M$ and there is no transition $\bar{x}^i \to x$,
then there is a trajectory from $\booltostr(x)$ to $\booltostr(\bar{x}^i)$ in the asynchronous dynamics of the $\T$-refinement of $M$.
\end{proposition}
\begin{proof}
Define $y^{\sigma}=\booltostr(x)+\varepsilon \sigma \mathrm{e}^i$ for all $\sigma = 0, \dots, m^i$, where $\varepsilon=\mathrm{sign}(\refinement(f, \T)_i(x)-x_i)$.
We have $y^0 = \booltostr(x)$ and $y^{m^i}=\booltostr(\bar{x}^i)$.
In addition, $\mvtobool^i(y^\sigma)_j=x_j$ for all $j \neq i$, and since $f_i(x)=f_i(\bar{x}^i)$ we get $\refinement(f, \T)_i(y^\sigma)=m^i \cdot f_i(x)$ for all $\sigma$, and there is a transition $y^\sigma\to y^{\sigma+1}$ for all $\sigma =0,\dots, m^i-1$.
\end{proof}

The interaction graph $G$ of a Boolean network $M=(V,f)$ can be endowed with a label function $S\colon E\to\P(\{-1,1\})$ that assigns signs to edges.
For an edge $(j,i)$ in $E$ and $s\in\{-1,1\}$, we have $s\in S((j,i))$ if there exists a state $x \in \B^V$ such that $(f_i(\bar{x}^j)-f_i(x))(\bar{x}^j_j-x_j)=s$.
\cref{prop:boolinstr} then gives the following corollary.

\begin{corollary}\label{cor:paths-ref}
Let $M=(V,f)$ be a Boolean network and suppose that the interaction graph of $f$ has no loops with negative sign.
If there is a path from $x$ to $y$ in the asynchronous dynamics,
then there is a path from $\booltostr(x)$ to $\booltostr(y)$ in the asynchronous dynamics of all single threshold refinements of $M$.
\end{corollary}

For some single threshold refinements of Boolean networks with negative loops in the interaction graph, the asynchronous dynamics can contain oscillations at intermediate
levels and fail to capture the Boolean dynamics.

\begin{example}
  Consider the Boolean network $(\{1,2\}, f)$ with $f(x_1,x_2)=(\bar{x}_1, x_1)$.
  The map $\T\colon \{1,2\}^2\to\mathbb{N}^*$ defined by $\T(1,1)=1$, $\T(1,2)=2$, $\T(2,1)=\T(2,2)=0$ is a threshold map for $f$.
  The associated $\T$-refinement is given by $\M=\{0,1,2\}\times\{0,1\}$,
  $\refinement(f, \T)_1(y_1,y_2)=2f_1(\ind(y_1\geq 1), 1)$, $\refinement(f, \T)_2(y_1,y_2)=f_2(\ind(y_1\geq 2), 1)$,
  so that $(0,0)$ and $(0,1)$ are mapped to $(2,0)$,
  $(1,0)$ and $(1,1)$ are mapped to $(0,0)$,
  and $(2,0)$ and $(2,1)$ are mapped to $(0,1)$.
  There is a transition from $(0,0)$ to $(1,0)$ in the Boolean asynchronous dynamics,
  but there is no trajectory from $\booltostr(0,0)=(0,0)$ to $\booltostr(1,0)=(2,0)$ in the multi-valued asynchronous dynamics.
\end{example}

\begin{definition}
Let $M = (V,f)$ be a Boolean network, $L \subseteq E$ a subset of its interactions, $M^L = (V \cup L, \extension(f,L))$ the associated
extension. Let $\T$ be a threshold map for $M$, with $\M$ the associated multi-valued space.
We define the mapping $\mvtobuf \colon \M \to \{0,1,\star\}^{V \cup L}$ as follows:
$$\mvtobuf(x)_i = \begin{cases}
  0 & \text{if } i \in V \text{ and } x_i = 0 \text{,} \\
  \star & \text{if } i \in V \text{ and }  0 < x_i < m^i \text{,} \\
  1 & \text{if } i \in V \text{ and } x_i = m^i \text{,} \\
  \ind(x_j \geq \T(j,k)) & \text{if } i = (j,k) \in L\text{.}
\end{cases}$$
for all $x \in \M$ and $i \in V \cup L$.
\end{definition}

If $x \in \B^V$ is a state of the Boolean network, then $\mvtobuf(\booltostr(x)) = \E(x)$.

\begin{proposition}
\label{prop:strinfull}
Let $M = (V,f)$ be a Boolean network, $(\M, \refinement(f, \T))$ the single threshold refinement of $M$ associated to a threshold map $\T$ and $(V \cup E, \extension(f, E))$ the full extension of $M$.
If there is a transition $x \to y$ in the asynchronous dynamics of $\refinement(f, \T)$, then for each state $z \in \mvtobuf(x)$ there is a geodesic from
$z$ to at least one state $z' \in \mvtobuf(y)$ in the asynchronous dynamics of $\extension(f, E)$.
\end{proposition}
\begin{proof}
Let $i$ be the only component such that $x_i \neq y_i$. We call $v = f_i(\mvtobool^i(x))$ the Boolean target value of $i$ at $\mvtobool^i(x)$.
We have $x_i \neq \refinement(f, \T)_i(x) = m^i \cdot v$.
Take a state $z \in \mvtobuf(x)$. For each regulator $j$ of $i$ we have $\pi^i(z)_j=z_{(i,j)}=\mvtobool^i(x)_j$, hence $\extension(f, E)_i(z) = f_i(\pi^i(z)) = f_i(\mvtobool^i(x)) = v$.

By definition of $\mvtobuf$, $\mvtobuf_i(y) \in \{v, \star \}$. Call $w \in \mvtobuf(y) \in \B^{V\cup E}$ the unique state such that $w_i = v$
and $\Delta(z,w) \subseteq \{i\} \cup \{(i, k)\ \mid \ k \in T(i)\}$.
We will show that there is a geodesic from $z$ to $w$.

As the extended network is a full extension, all targets of $i$ in the interaction graph of $\extension(f, E)$ are in $E$.
Let $U = \Delta(z,w) \setminus \{i\} = \Delta(z,w) \cap E$ be the set of targets of $i$ that differ in $w$ and $z$.
For each $e=(i,k) \in U$, we have $\ind(x_i \geq \T(i,k)) = z_e \neq w_e = \ind(y_i \geq \T(i,k)) = v$.

If $z_i = v$ then there is a geodesic from $z$ to $w$ that consists in updating all components of $U$
(this is possible in any order).
If $z_i \neq v$ then since $\extension(f, E)_i(z) = v$ there is a transition $z \to \bar{z}^i$, followed by a similar geodesic from $\bar{z}^i$ to $w$.
\end{proof}

\begin{corollary}
\label{cor:strinfull}
Consider a Boolean network $(V,f)$ and $x, y$ Boolean states.
If there exists a threshold map $\T$ such that $\booltostr(y)$ is reachable from $\booltostr(x)$ in the asynchronous dynamics of $\refinement(f, \T)$, then $y$ is $E$-reachable from $x$.
\end{corollary}

Note that in \cref{prop:strinfull} and \cref{cor:strinfull}, we only considered the full extension. Whether the conclusions hold for any cuttable extension remains an open question. 

\begin{figure}
  \begin{center}
    \includegraphics[width=\figwidth]{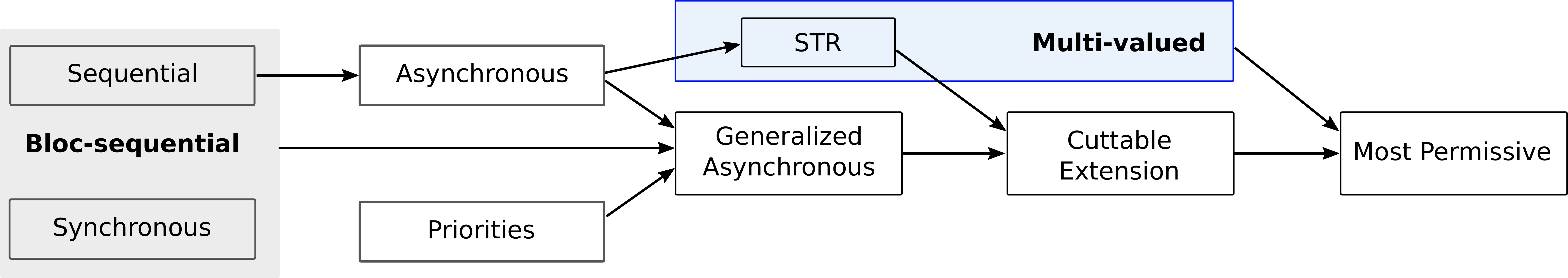}
  \end{center}
  \caption{\label{fig:reach}Reachability properties across updating semantics.
  Boxes represent updating semantics and arrows between them indicate that the target semantics is an over-approximation of the source semantics.
  The gray area on the left groups classical deterministic semantics, while all others are non-deterministic.
  STR stands for single threshold refinement (\cref{def:str}),
  and the blue area denotes the asynchronous semantics of all multi-valued refinements.
  }
\end{figure}

\section{Discussion}\label{sec:discussion}

To reflect the lack of kinetic knowledge often associated with biological networks, the classical asynchronous semantics explores
all possible alternative trajectories where a single component is updated in each transition. The generalized asynchronous
semantics accounts for possible partial or total synchronism in updates.
The binary nature of activity levels on the other hand implies that a change of the activity level of a single component simultaneously affects all its target components.
In many networks, the effect of a component on different targets involves different mechanisms with their own kinetics and even
sometimes different implicit intermediates. In case of competition (such as the inconsistent feedback loop in \cref{fig:trajs}~a),
the classical semantics then fail to capture some plausible behaviours.
Multi-valued networks could be used to define separate thresholds for different targets, but would require either additional knowledge for
all interactions or the identification of some key interactions that would benefit from a refinement.
The most permissive semantics uses transitory states to address this issue and reproduce the behaviour of all multi-valued refinements,
but also introduces undesired non monotonic behaviours. For example, a component in the increasing state can act in succession as
inactive, then active, then inactive again for one of its targets as illustrated in \cref{fig:trajs} b).
While such behaviours could be interpreted as stochastic effects in the neighbourhood of an activation threshold, they can often be
considered as artefacts.
Here, we focused on single threshold refinements, a small subset of multi-valued refinements that enable threshold separation while preserving
the original Boolean functions (thus without introducing non monotonic behaviours). The extension of individual interactions with linear
components can be used to emulate such refinements in absence of knowledge on the threshold values and within the established
framework of asynchronous Boolean networks.

As a tool to study asynchronous trajectories we introduced implicant maps representing dependencies and conflicts controlling the possible change of value of the components
compared to a specific initial state. These implicant maps correspond to classes of subgraphs in the implicant graph used for the identification
of trap spaces \citep[stable motifs, see][]{Zanudo2013stablemotifs} or equivalently in the Petri net unfolding of the Boolean network~\citep{Chaouiya2011pn}.
We say that an implicant map is weakly consistent if it describes a set of satisfiable (complete and non-circular) dependencies.
In absence of any weakly consistent map containing a given component, we know that there is no trajectory (in any semantics) in which the value of
this component can be modified. This strong requirement is consistent with our observation that the maximal weakly consistent maps correspond to the
smallest trap spaces containing the initial state.
This weak consistency solely relies on dependencies and ignores the competition between components. In permissive trajectories this limitation is
ignored and all components included in a weakly consistent map can be updated in a geodesic (following a partial order defined by the dependencies).
However these competitions can play a role in asynchronous trajectories, where some of these components can only be updated after much longer
trajectories, if ever. A weakly consistent implicant map is strongly consistent in absence of competition between its components. This stronger
consistency property is both necessary and sufficient for the existence of asynchronous geodesics.

As the direct requirements and competitions described by implicant maps are associated to interactions in the regulatory graph,
the consistency constraints correspond to undirected cycles in the interaction graph. We further observed that a linear component
mirroring its unique regulator in the initial state can be used to relax such competitions.
This led us to study the dynamical properties of cuttable networks, a structural class of Boolean networks in which a set of linear
components cover all feedback loops and paths from any component with multiple targets to any component with multiple regulators.
Our observations suggest that these two structural conditions correspond to different types of competitions.
On one hand, the linear extension of feedback loops seems to be associated to synchronized update of multiple components, as illustrated in \cref{fig:trajs} d).
It is thus required and could be sufficient to reproduce the generalized asynchronous trajectories.
On the other hand, the linear extension of paths connecting a component with multiple targets to a component with multiple regulators could
be related to threshold separation in feedforward loops. We observed strong similarities between the trajectories recovered through the extension of
feedforward loops and in single threshold refinements as illustrated in \cref{fig:trajs} a,c).
These two associations are consistent with the fact that the extended dynamics reproduces the reachability properties obtained in both the
generalized asynchronous and all single threshold refinements. Further work is needed to clarify the role of feedback loops, feedforward loops,
and other paths from components with multiple targets to components with multiple regulators in the dynamical properties of cuttable networks to
elucidate whether the structural conditions for linear cuts could then be further generalized.

We have implemented the linear extension of Boolean networks in the bioLQM software~\citep{Naldi2018bioLQM}, enabling
the use of the extended semantics in existing software tools supporting the classical asynchronous semantics.
Note that efficient analysis based on trap spaces does not require this explicit extension and can be performed directly on
the original Boolean networks using existing implementations of trap spaces identification in PyBoolNet~\citep{klarner2017pyboolnet}
or BioLQM.

As shown by \citet{klarner2014trapspaces}, prime implicants provide a compact and complete representation of the implicant graph enabling
the identification of sets of implicants that cooperatively define a trap space as the solutions of a constraint solving problem. We plan
to adapt this approach to the identification of implicant maps with the desired consistency level. The identification of strongly-consistent
maps can be used as a proof of reachability in the asynchronous semantics, while the identification of weakly consistent maps
can be used to pinpoint specific competitions that need to be relaxed to enable this reachability. Beyond the general question of
reachability, this approach would provide valuable hints to assess the biological relevance of the corresponding extended trajectories.
Note that this type of reasoning can only be used to formally validate a reachability property: if the competitions can not be realistically
relaxed, then more complex trajectories to the target of interest may still exist.

\section{Conclusion}\label{sec:conclusion}

In this paper we study the reachability properties of dynamical Boolean networks, and in particular the reachability of a subspace
from a specific initial state. This question is known to be PSPACE-complete in the classical asynchronous semantics, 
however abstract interpretation approaches provide efficient solutions in some cases \citep{Pauleve2012abstract,pauleve2020most}.
Furthermore, this problem is polynomial for monotonic networks in the recently proposed most permissive semantics~\citep{pauleve2020most}.
This novel semantics extends the classical asynchronous semantics by adding intermediate activity levels explicitly accounting for the
absence of information on the regulation thresholds. This approach enables the simulation of relevant behaviours missed by the standard asynchronous dynamics.
The most permissive semantics can, on the other hand, also introduce some artefactual behaviours and should thus be considered as an over-approximation.
This work starts with the characterisation of different structural conditions for individual transitions in asynchronous and permissive
trajectories and leads to the identification of a class of Boolean networks and initial states for which these semantics have the
same geodesics.
These networks have a simple structural characterization: they are networks whose interaction graph admits a \emph{linear cut}.
We could show that trap spaces \citep[also called stable motifs or symbolic steady states, see][]{Zanudo2013stablemotifs,klarner2014trapspaces}
always provide a precise characterization of all attractors in cuttable networks, and that their reachability solely depends on the minimal
trap space containing the initial state. These results are strong improvements compared to the general case where trap spaces lack such formal
guarantees, even if they are often considered as good estimators in practice.
These results are similar to the properties of the most permissive dynamics but here they do not rely on intermediate activity levels that
could induce known artefactual behaviours.

We then proposed an extended semantics based on linear extensions of Boolean networks.
This type of extension can be interpreted as the explicit representation of hidden delays or threshold effects,
and thus carries a natural biological justification. As trap spaces of the original network are also trap
spaces of their extensions, the properties of cuttable networks (reachability of trap spaces and configuration of attractors) can then be
applied directly to any Boolean network without explicitly constructing a cuttable extension.
The reachability properties of this extended semantics provide an interesting middle ground between the asynchronous semantics and the most permissive semantics,
as it recovers realistic trajectories missing in the former and excludes some artefactual behaviours of the latter (see~\cref{fig:reach}).
The reachability of trap spaces in the cuttable extension semantics has the same polynomial complexity as in the most permissive; however, the reachability
of transient subspaces remains to be investigated. It is currently unclear if all permissive trajectories which are not captured by this new semantics are
associated to non-monotonicity (and could be considered as artefacts) or if some relevant trajectories (to transient states) might also missing.
Similarly, while the most permissive semantics capture all possible behaviours of multi-valued refinements,
the ability of our extended semantics to reproduce behaviours emerging in multi-valued refinements has been only partially explored.
We have shown that refinements that rely on a unique threshold per regulation can be captured by full extensions;
however this condition does not fully characterized the emerging behaviours.

The strength of Boolean networks lies in their simple, parameter-free formulation.
However, their ability to deal with lack of detailed kinetic information is also at the core of their intrinsic limitations.
Although the parameter uncertainty can partially be encoded by resorting to non-deterministic semantics,
many potential fine-grained behaviours that depend on specific parameter scenarios are inevitably inaccessible when relying to logical rules alone.
The most permissive semantics provide an important step to ensure that all possible parameters are indeed captured, and can thus be
used to formally rule out reachability properties which are structurally impossible for any set of parameters. However, it also
increases the number of artefactual trajectories in the system. Implicant maps provide the groundwork to formally identify trajectories
which remain realistic for any set of parameters or for parameters matching well-characterized conditions. These maps can be constructed
for direct trajectories (geodesics) in the permissive or extended semantics as shown here and could be naturally extended to trajectories
where all components are updated at most twice, which can be required for the reachability of some trap spaces. However, it would not
scale to arbitrarily complex trajectories, which remain in a gray area. We could imagine combining these approaches to annotate any
reachability property as formally impossible, unlikely, realistic or formally guaranteed.

\paragraph*{Authors' contributions}

AN and ET conceived and developed the project.
All authors expanded and formulated the theory, contributed to the manuscript, read and approved the final manuscript.

\paragraph*{Acknowledgments}

We thank Heike Siebert for insightful discussions.

\paragraph*{Funding}

AN was supported by the Deutsche Forschungsgemeinschaft (GRK 1772) and French National Research Agency (ANR-MOST project ANR-16-CE18-0029 "BIOPSY").
AR was supported by the French National Research Agency (Young Researcher project ANR-18-CE40-0002-01 "FANs").

\paragraph*{Competing interests}

The authors have no relevant financial or non-financial interests to disclose.

\end{document}